\documentstyle[aps,epsf,psfig,floats,twocolumn]{revtex}

\def\vr{{\bf{r}}}
\def\vt{{\bf{t}}}

\def\dd{{\rm d}} 
\def\writhe{\chi_W}
\def\chif{{\writhe^F}} 
\def\chic{{\writhe^C}}

\begin{document}
{\bf Comment on ``Elasticity Model of a Supercoiled DNA Molecule''}

Bouchiat and M\'ezard\cite{MezardPRL} have proposed a ``worm-like rod
chain model'' to describe recent experiments on DNA
mechanics\cite{ABCS}.  Their model can be solved using an analogy with
quantum mechanics, thanks to the {\em local} nature of the effective
Hamiltonian.  The letter concluded that this model has exceptional
physical behavior: The continuum limit is singular and linear response
theory can not be used in the absence of a cut-off to calculate the
experimental curves: For example the curves of extension versus
torsion flatten in the continuum limit.

The writhe, $W\!r=\chic/2\pi$, of a chain is given by the
C\u{a}lug\u{a}reanu-White formula\cite{Calugareanu}
\begin{equation}
\chic=\frac{1}{2}
\int\!\!\!\int_0^L \frac{\vr(s)-\vr(s')}
{{|\vr(s)-\vr(s')|}^3}\cdot
\frac{\dd\vr(s)}{\dd s}\times
\frac{\dd\vr(s')}{\dd s}\,\dd s\,\dd s',
\label{Gauss}
\end{equation}
where $\vr(s)=\int_0^s \vt(s')\,\dd s'$ is the position in space
and~$L$ is the length of the chain.  Fuller showed that if we
represent the tangent vector~$\vt$ by a point on the unit sphere the
area, $\chif$, enclosed by~$\vt(s)$ on the sphere\cite{Fuller}, is in
certain cases equal to~$\chic$.  The general relation between Fuller's
and C\u{a}lug\u{a}reanu's formulas is expressed by:
\begin{equation}
\chif\equiv\chic {\rm\ mod.\ } 4\pi.
\label{modulo}
\end{equation}
Fuller's result, expressed as a local integral, was used in
\cite{MezardPRL}.  The authors justify the use of $\chif$ to interpret
experiments on {\sl open} chains with a topological modification of
the sphere \cite{MezardEPJE} and find that the probability
distribution $p(\chif)$ varies as $1/\left(\small\chif\right)^2$ for
large~$\chif$ due to winding of trajectories about the south pole of
the sphere.

We have performed a simulations to directly compare the distribution
of writhe implied by each formulation: We numerically generate a large
equilibrated ensemble of semiflexible chains of eight times the
persistence length, made of $8N$ straight links with $\vt(0)$ and
$\vt(L)$ parallel.  This assures that the path of~$\vt(s)$ on the
sphere is a closed loop. In order to use the C\u{a}lug\u{a}reanu-White
formula, the chain must also be closed in real space.  To do this we
extend the chain at each end with a long straight segment in the
directions $-\vt(0)$ and $\vt(L)$ and then join the two extremities
with an arc of a circle; the contribution from the exterior is
important in order to preserve the modulo equality eq.~(\ref{modulo});
it is this augmented version of eq.~(1) which determins the rotation
angle measured in experiments on open chains.  We calculate $\chif$
and $\chic$ for each chain from our ensemble and plot the integrated
probability $P(\writhe)=\int^{\writhe}_{-\infty}p(\chi) \dd \chi$.
For the Cauchy distribution, one thus obtains $\chif\sim\frac{1}{1-P}$
for large values of~$\chif$.
 
\begin{figure}[ht]
\begin{center}
  \epsfxsize=8cm \epsfbox{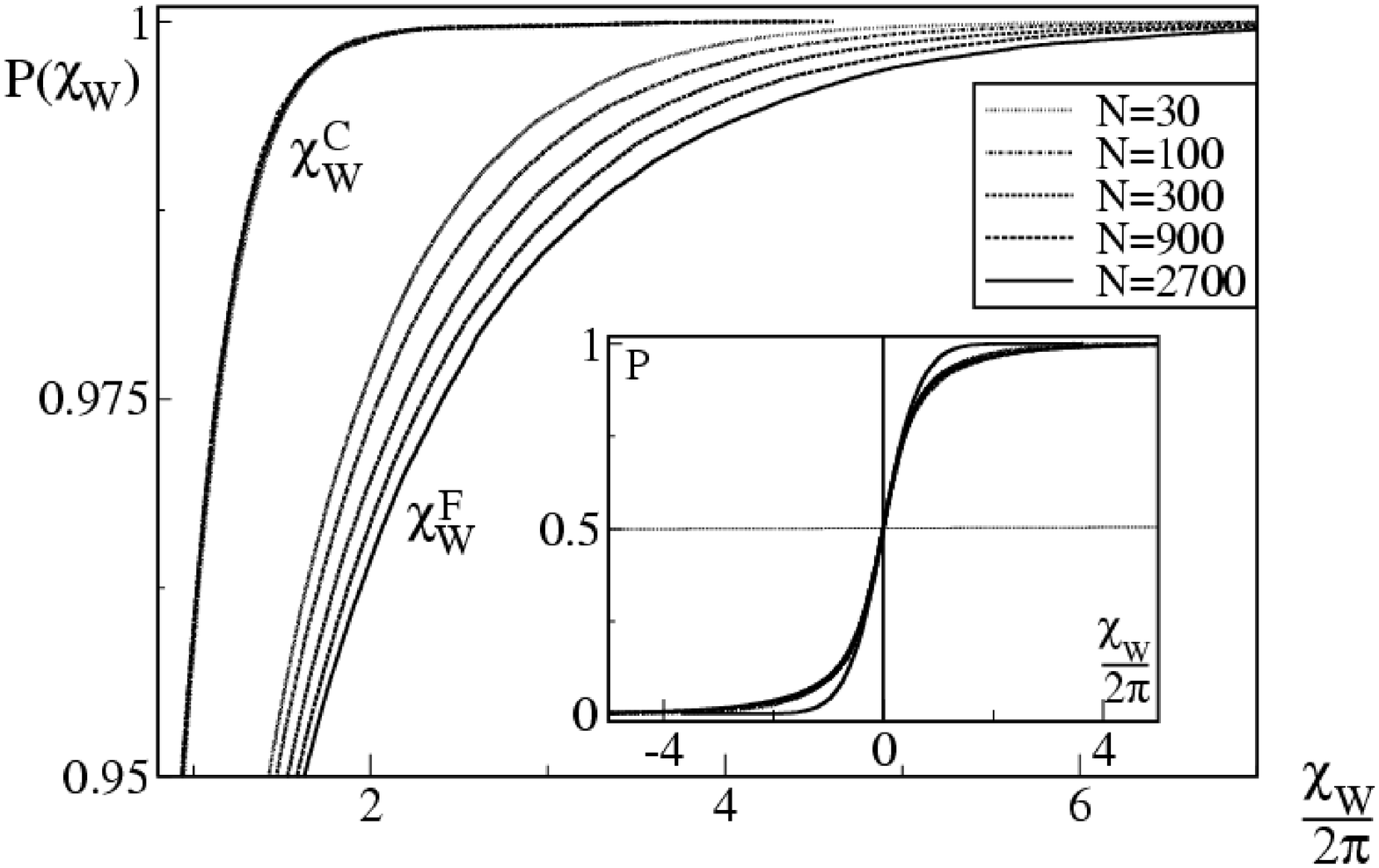} \vspace{0.15cm}
\caption[]{
  Inset: The integrated writhe distribution function $P(\writhe)$
  computed with two different formulations of the writhe.
  $\chic$ not computed for $N=2700$ due to computer limitations.  The
  main figure is a zoom of the top right corner of the inset. One can
  compare the continuous evolution of~$\chif$ as $N$ grows with the
  stability of~$\chic$ in the same limit. An asymptote in
  $\chif\sim\frac{1}{1-P}$ observed in this figure corresponds to the
  Cauchy distribution.  The singularity of~$\chic$ is clearly weaker
  than~$\frac{1}{1-P}$.  Statistical fluctuations of the curves are
  smaller than their width.}{}
\label{proba}
\vspace{-0.8cm}
\end{center}
\end{figure}

Results are displayed in fig.~\ref{proba}.  The curves of $P$ as a
function of $\chif$ indeed depend on discretization. We have
additionally verified that the curves of $\chif$ scaled by the factor
$\sqrt{\ln N}$ converge very accurately to a master curve as predicted
in \cite{MezardPRL}.  On the other hand, $\chic$ converges to a
limiting curve as~$N$ grows.  We find no evidence of divergence
of~$\chic$ as~$\frac{1}{1-P}$ nor evidence of scaling with~$\sqrt{\ln
  N}$.  From fig.~\ref{proba} we see that even for $N$ small, the
Fuller formulation strongly overestimates the statistical weight of
large writhe fluctuations.

To conclude we believe that torsional fluctuations of a semiflexible
chains are not divergent when going to the continum limit. A DNA chain
has a persistence length of order 50 times its radius; it is in a
regime where eq.~(\ref{Gauss}) has reached the continuum limit to a
very good approximation.

\medskip
{\noindent\small V.~Rossetto and A.C.~Maggs,\\
  PCT, ESPCI CNRS, 10 rue Vauquelin, 75005, Paris.}

\end{document}